\def\tcar{\futurelet\next\testnextcar}
\def\testnextcar{\ifhmode\ifcat\next.\else\ \fi\fi}
\def\kms{km~s$^{-1}$\tcar}

\def\etal{{et~al.}\ }

\def\QSO1038{{Q~1038$-$2712}}
\def\qso2343{Q~2343$+$125}
\def\pg2302{PG~2302$+$029}
\def\Lya{\hbox{Ly-$\alpha$~}}
\def\Lyb{Ly-$\beta$~}
\def\Lyg{Ly-$\gamma$~}
\def\cycle1{Cycle~1}
\def\eg{{\it{e.g.~}}}

\documentstyle[12pt,aaspp4,flushrt]{article}
\begin{document}

\overfullrule=0pt

\title{THE HST QUASAR ABSORPTION LINE KEY PROJECT XII: The Unusual
Absorption Line System in the Spectrum of PG~2302$+$029 
--- Ejected or Intervening?\altaffilmark{1}
\altaffiltext{1}{Based on observations with the
NASA/ESA {\it Hubble Space Telescope}, obtained at the Space Telescope 
Science Institute, which is operated by the Association of
Universities for Research in Astronomy, Inc., under NASA contract
NAS5-26555.}}

\author{
B.~T.~Jannuzi,\altaffilmark{2} \altaffiltext{2}{National Optical
Astronomy Observatories, P.O. Box 26732, Tucson, AZ~85719,
email:~jannuzi@noao.edu}
G.~F.~Hartig,\altaffilmark{3}\altaffiltext{3}{Space Telescope Science 
Institute, 3700~San Martin Drive, Baltimore, MD~21218} 
S.~Kirhakos,\altaffilmark{4}\altaffiltext{4}{Institute for Advanced 
Study, School of Natural Sciences, Olden Lane, Princeton, NJ~08540} 
W.~ L.~W.~Sargent,\altaffilmark{5}\altaffiltext{5}{Robinson Laboratory 
105-24, California Institute of Technology, Pasadena, CA 91125}
D.~A.~Turnshek,\altaffilmark{6}\altaffiltext{6}{Department of
Physics \& Astronomy, University of Pittsburgh, Pittsburgh, PA 15260}
R.~J.~Weymann,\altaffilmark{7}\altaffiltext{7}{The Observatories of
the Carnegie Institution of Washington, 813 Santa Barbara Street,
Pasadena, CA 91101} 
J.~N.~Bahcall,\altaffilmark{4}
J.~Bergeron,\altaffilmark{8}\altaffiltext{8}{Institut d'Astrophysique, 
98 bis Boulevard Arago, F-75014 Paris, France} 
A.~Boksenberg,\altaffilmark{9}\altaffiltext{9}{Institute of Astronomy,
University of Cambridge, Madingley Road, Cambridge CB3 OHA, UK}
B.~D.~Savage,\altaffilmark{10}\altaffiltext{10}{Department of Astronomy, 
University of Wisconsin, 475 N. Charter Street, Madison, WI 53706} 
D.~P.~Schneider,\altaffilmark{11}\altaffiltext{11}{Department of 
Astronomy, The Pennsylvania State University, University Park, 
PA 16802}
\& A.~M.~Wolfe\altaffilmark{12}\altaffiltext{12}{Center for
Astrophysics \& Space Sciences, C011, University of California San
Diego, La Jolla, CA 92093} }

\begin{abstract}

We report the discovery of a high-ionization broad absorption line
system at a redshift of $z_{\rm{abs}}=0.695$ in the spectrum of the
$z_{\rm{em}}=1.052$ radio-quiet quasar PG~2302$+$039.  Broad
absorption with FWHM from 3,000~to~5,000~\kms is detected from C~IV,
N~V, and O~VI in HST Faint Object Spectrograph spectra of the quasar.
A narrow line system (FWHM $\sim250$~\kms) at $z_{\rm{abs}}=0.7016$ is
resolved from the broad blend and includes absorption by \Lya and the
C~IV, N~V, and O~VI doublets.  No absorption by low-ionization metal
species (\eg Si~II and Mg~II) is detected in the HST or ground-based
spectra for either the broad or the narrow system. The centroids of
the broad system lines are displaced by $\sim56,000$~\kms to the blue
of the quasar's broad emission lines. The reddest extent of the broad
line absorption is more than 50,000~\kms from the quasar.  The
properties of this system are unprecedented whether it is an
intervening or an ejected system.

\end{abstract}

\section{Introduction}
\label{sec-intro}

Extremely broad high-ionization absorption complexes with velocity
extents of 2,000 to 25,000~\kms occur in approximately 10\% of
radio-quiet quasi-stellar objects (QSOs) (Weymann \etal 1991).  This
absorption is almost certainly produced by material ejected from the
source producing the observed emission.  Such objects are Broad
Absorption Line QSOs (BALQSOs; see Weymann \etal 1991 and Turnshek
1995 for reviews). Broad absorption features produced by intervening
gaseous material associated with clusters of galaxies were anticipated
by Bahcall \& Salpeter (1965), but only a small number of such
features have been convincingly demonstrated.  One dramatic example is
the extensive complex of absorption near $z_{\rm{abs}}=2$ in the
spectra of two quasars (Q~1037$-$2704, $z_{\rm{em}}=2.193$, and
Q~1038$-$2712, $z_{\rm{em}}=2.33$, the ``Tololo Pair'') separated by
17.9$'$ on the sky (Jakobsen \etal 1986). It is very probable that the
extensive absorption observed in these quasars (and in additional
quasars subsequently observed in the same field) is produced by
gaseous material associated with a supercluster of galaxies (Sargent
\& Steidel 1987; Jakobsen \& Perryman 1992; Dinshaw \& Impey 1996).
Broad absorption features in quasar spectra can evidently be a tracer
both of large-scale structure in the Universe and of material
intrinsic to the QSOs.

In this {\it Letter} we report the discovery of a very broad
high-ionization absorption complex at a redshift of $\sim0.695$~ in
the spectrum of \pg2302 (\hbox{$\alpha \ = \ 23^{\rm h} \ 04^{\rm m} \
45.0^{\rm s}$}, \hbox{$\delta \ = \ +03^{\circ} \ 11' \ 46''$}, J2000,
Schneider \etal 1992). The quasar has an emission line redshift of
$z=1.052$ (Steidel \& Sargent 1991) and is radio-quiet (Kellerman
\etal 1989).

\begin{table}[h]
\small
\caption{HST/FOS Observations of PG2302+029}
\begin{tabular*}{6.5in}{l@{\extracolsep{\fill}}clcccc}
\noalign{\medskip}
\hline\hline
\noalign{\medskip}
{UT Date}&{Exposure}
&Detector/&Apt.
&FWHM$^{\rm{a}}$&S/N$^{\rm b}$ & Shift$^{\rm c}$ \\ 
&\multicolumn{1}{c}{Time
(s)}&\multicolumn{1}{c}{Grating}&\multicolumn{1}{c}{\arcsec}&
\multicolumn{1}{c}{(\AA)}&&\multicolumn{1}{c}{(\AA)}\\
\noalign{\medskip}
\hline
\noalign{\medskip}
1994~Oct~05$^{\rm d}$& 4422 &Red/G190H & 0.26 &1.33& 30 &  0.85   \nl
1994~Aug~02$^{\rm e}$& 1400 &Red/G270H & 0.86 &1.97& 50 &  1.14   \nl
1992~Jul~28$^{\rm f}$& 7920 &Blue/G270H & 4.30 &$\sim 4.0$&80 & ---   \nl
\noalign{\medskip\hrule\medskip}
\noalign{\vbox{\hsize=6.5in\baselineskip=14pt
$^{\rm a}$ { The size in ~\AA~ of a spectral resolution element. }\hfil\break
$^{\rm b}$ { Signal-to-noise ratio per resolution element in
the continuum.}\hfil\break
$^{\rm c}$ { The wavelength zero point offsets which were
applied to the observed G190H and 1994 G270H spectra in order to place
them in a frame with Galactic interstellar medium lines at rest (see
Schneider \etal 1993 for discussion).}\hfil\break
$^{\rm d}$ { Spectrum obtained by Key Project, Bahcall PI.}\hfil\break
$^{\rm e}$ { Spectrum obtained by GTO program, Burbidge PI.}\hfil\break
$^{\rm f}$ { Spectropolarimetry data obtained with the
pre-COSTAR FOS; published by Impey {\it et~al} 1995.}
}}
\end{tabular*}
\end{table}

\section{Observations} 
\label{sec-obs}

HST Faint Object Spectrograph (FOS; description by Ford \& Hartig
1990) spectra obtained using the G190H (covering $\sim1600-2300$~\AA)
and G270H ($2250-3250$~\AA) gratings provide the new data presented in
this {\it Letter}.  Characteristics of the observations are presented
in Table~1 and the spectra are displayed in Figure~1.  We obtained the
G190H
spectrum as part of the HST quasar absorption line survey, an
HST Key Project during Cycles $1-4$~ (Bahcall \etal 1993).  The
spectrum was processed using IRAF as described by Schneider~\etal
1993.  Burbidge \etal obtained a G270H spectrum of \pg2302 in 1994.
In 1992 Impey \etal (1995) made spectropolarimetry observations of
\pg2302 using the G270H grating, the pre-COSTAR FOS, and a large
aperture ($4.3''$), resulting in significantly lower spectral
resolution in their spectrum than either the G190H or 1994 G270H
spectra. Both of the G270H spectra were extracted from the HST
archive, making use of the standard STScI processing of the data.

The spectra were searched for absorption lines which were fit with
Gaussians using the Key Project software described in Schneider~\etal
1993.  The 1992 G270H spectrum was only used to check for variability
of spectral features (no significant changes were found). Measurements
and identifications of all the absorption features will be included in
the next Key Project catalogue paper.  The focus of this {\it Letter}
are the broad absorption features and their associated narrow
components.  Measurements of these lines are presented in Table~2.
The continuum fit was subjectively adjusted in the regions of the
broad features and the separation of broad and narrow features near
2100~\AA~ was done interactively.

\begin{table}[t]
\def\Lye{Ly-$\epsilon$}
\def\Lya{Ly-$\alpha$}
\def\Lyb{Ly-$\beta$}
\def\Lyd{Ly-$\delta$}
\def\Lyg{Ly-$\gamma$}
$$\vbox{\small\baselineskip=14pt
 \halign{#\hfil\tabskip=1em plus.5em minus.5em&
\hfil#&
\hfil#&
\hfil#&
\hfil#&
\hfil#&
#\hfil&
\hfil#&
\hfil$#$&
#\hfil\tabskip=0pt\cr
\multispan{10}{Table 2:\ PG 2302$+$029  ($z$= 1.052) Selected 
Redshift Systems\hfill}\cr
\noalign{\medskip\hrule\smallskip\hrule\medskip}
\hfil $\lambda_{\rm obs}$\hfil&\hfil $\sigma(\lambda)$\hfil&
\hfil $W_{\rm obs}$\hfil&
\hfil $\sigma(W)$\hfil&
\hfil SL$^{\rm{a}}$\hfil&
\hfil FWHM\hfil&
\multispan2{\hfil Line ID\hfil}&\hfil \Delta\lambda\hfil&
\hfil $z_{\rm obs}$\hfil\cr
\noalign{\vglue-12pt}
&&&&&&\multispan2{\hrulefill}\cr
~~(\AA)&(\AA)&(\AA)&(\AA)&&(\AA)& ION &(\AA)~~&({\rm\AA})&\cr
\noalign{\medskip\hrule\medskip}
 1750.49& 0.07& 12.56& 0.65& 144.88& 28.02& O~VI Bl & 1034.79& -3.58& 0.695~\cr
 1773.86& 0.44& 1.13& 0.27&  17.71& 6.05& O~VI? Bl & 1034.79& 19.79&  0.695~\cr
 2103.24& 0.55& 5.67& 0.25& 147.11& 24.23& N~V Bl &  1240.81& 0.00&  0.695~\cr
 2626.52& 0.83& 6.34& 0.43& 130.00& 26.20& C~IV Bl & 1549.49& 0.00&  0.695~\cr
 1661.55& 0.15& 1.75& 0.17& 17.10& 3.29&  C~III? Bl& 977.03&  -0.95&  0.7016\cr
 1755.75& 0.15& 0.54& 0.11& 6.83& 2.08&  O VI &  1031.95&  -0.18&  0.7016\cr
 1765.20& 0.14& 0.66& 0.14& 8.26& 1.83&  O VI &  1037.63&  -0.39&  0.7016\cr
 2068.50& 0.06& 1.06& 0.09&   25.40& 1.75&  \Lya &  1215.67&  -0.03&  0.7016\cr
 2107.92& 0.16& 0.23& 0.06& 7.01& 1.65&  N V & 1238.82&  -0.01&  0.7016\cr
 2113.85& 0.22& 0.35& 0.08& 10.59& 2.51&  N V & 1242.80&  -0.84&  0.7016\cr
 2634.53& 0.12& 0.66& 0.10& 13.60& 2.85&  C IV &  1548.20&   0.18&  0.7016\cr
 2638.72& 0.12& 0.45& 0.09& 9.30& 2.06&  C IV &  1550.77&   0.00&  0.7016\cr
%
%
%
\noalign{\medskip\hrule\medskip}
\noalign{\vbox{\hsize=5.75in
$^{\rm a}$\ SL$=$ the significance level of the line; definded to
be the measured equivalent width ($W$) divided by the one sigma
uncertainty in the measurement of an unresolved line at the same
wavelength. See Schneider \etal 1993 for complete discussion.}}
}}$$
\end{table}

The narrow absorption lines were identified using the algorithmic line
identification software described by Bahcall \etal (1996).  Among the
identifications is a high-ionization $z_{\rm{abs}}=0.7016$ system
including \Lya, the C~IV, N~V, and O~VI doublets, and a possible
detection of C~III (the line is located in a low signal-to-noise
region of the G190H spectrum).  Fe~III, if present, is blended with
other lines.  The lines in this system are marginally resolved at the
resolution of our observations ($\sim250$~\kms).

In addition to the narrow features, the UV spectrum of \pg2302
includes three broad features centered at 2626.52, 2103.24, and
1750.49~\AA~ (blended with 1773.86~\AA).  Impey \etal (1996) noted the
2626~\AA~ feature, but did not offer an identification.  The G190H
spectrum revealed two broad absorption features which, when combined
with the line in the G270H spectra, allowed us to identify these
features as broad (or heavily blended) C~IV, N~V, and O~VI at
$z\sim0.695$.  In Figure~2 we display the broad and narrow systems
near $z=0.7$~ on a velocity scale zeroed on the center of the broad
C~IV absorption.  Unlike the narrower system at $z=0.7016$, we
identify no \Lya absorption associated with the broad system.  The
broad features have widths (FWHM) ranging from $\sim3,000$~\kms for
C~IV and N~V to $\sim5,000$~\kms for O~VI.

We can place limits on the amount of absorption that might be present
from broad components of \Lya, C~III (977.03~\AA), Si~III
(1206.51~\AA), and Si~IV (1393.76, 1402.77~\AA).  However, our limits
are very uncertain since the detection of weak, broad absorption is
very sensitive to the placement of the continuum. Placement of the
continuum is complicated by the broad emission lines from the quasar
and the low signal-to-noise ratio in some regions of the spectra.
Assuming that additional broad components of the $z\sim0.695$ system
would have widths comparable to those already observed, we place the
following $3\sigma$~ limits on the equivalent widths of broad
absorption: \Lya$< 6$~\AA, C~III$< 5$~\AA, Si~III$< 5$~\AA, and
Si~IV$<3$~\AA. No low-ionization lines associated with either the
broad or narrow systems described above (e.g. from Si~II, S~II, Fe~II,
C~II, O~I, Al~II, N~I, Mg~II, or Mg~I) are present in the FOS (no
unresolved lines with $W>0.4$~\AA) or optical spectra (Petitjean \&
Bergeron 1990; Steidel \& Sargent 1992).

\section{Discussion}
\label{sec-discuss}

We have considered three possible causes for the broad complex of
high-ionization absorption observed at $z_{\rm{abs}}\sim0.7$ in the
spectrum of \pg2302. These are: 1.)~Absorption by material ejected
from the QSO with an extremely high ejection velocity (56,000~\kms),
but no low velocity absorbing gas along the line of sight.
2.)~Absorption by material in the galaxies or intergalactic medium of
a cluster or supercluster of galaxies.  3.)~Absorption by gas in a
supernovae remnant in a galaxy at a redshift of about $0.7$.  With the
current observations we can not make a definite choice between these
alternatives. The third suggestion might produce a broad velocity
profile similar to what we observe. The narrow high-ionization system
that is resolved from the broad system might then be a galaxy
associated with the host galaxy of the supernovae (the centers of the
broad system's lines and of the detected narrow system are more than
1,500~\kms apart) or a chance coincidence.  In the remainder of this
section we consider the other two suggested causes of the broad
high-ionization system.


\pg2302 and its broad  absorption system have several similarities to
BALQSOs.  Wampler (1986) demonstrated that the spectrum of \pg2302,
particularly its very strong Fe~II emission, is remarkably similar to
the well known BALQSO PG~1700$+$518.  Typical of BALQSO absorption
systems, the \pg2302 $z=0.695$ system is composed of high-ionization
species and covers a large spread in velocity (more than
3,000~\kms). BALQSOs usually also have detected Si~IV and \Lya
absorption, but this is not always the case. When these lines are
present they are generally weaker than N~V and O~VI.  There are other
unusual features about the \pg2302 absorption if it is produced by
ejected material.  First, the absorption troughs of most BALQSOs begin
at or close to the systemic velocity of the broad emission line region
(BELR) and extend blueward to higher velocities, generally stopping by
25,000~\kms (see Figure 9 in Korista \etal 1993).  Korista \etal
(1993) demonstrate that the typical BALQSO absorption velocity
profiles are asymmetric with the strongest absorption at the lowest
ejection velocities, decreasing in strength with increasing blueshift
from the BELR.  In contrast, the \pg2302 broad system is detached from
the velocity of the BELR by at least 50,000~\kms and each component is
well fit by a Gaussian (i.e. a symmetric function).

While the \pg2302 broad system is not typical of BALQSOs, it does
share characteristics with a system that is produced by material
ejected from a QSO.  Sargent, Boksenberg, \& Steidel (1988) identified
a broad feature in the spectrum of \qso2343 (FWHM$\sim1,000$~\kms, but
not broad enough for \qso2343 to be considered a BALQSO) which they
interpreted as a detached C~IV doublet ($z_{\rm{abs}}=2.24$) ejected
at a velocity of 23,000~\kms from the QSO ($z_{\rm{em}}=2.515$).  As
is the case for \pg2302, there are no broad absorption features close
to the emission redshift of \qso2343.  Keck HIRES spectra analyzed by
Hamann, Barlow, \& Junkkarinen (1996) reveal dramatic variations over
a period of months in the strength of the \qso2343 broad absorption
features, confirming that the absorption is produced by ejected
material.  Like the \pg2302 broad system, the \qso2343 system lacks
any associated low-ionization absorption (\Lya would be blended with
the high redshift \Lya forest, but Hamann \etal indicate that any
\Lya absorption is still weak).  While the observed broad system in
\pg2302 is significantly different from any system identified in a
BALQSO, it is similar to the system observed in \qso2343 and might be
produced by material ejected from a QSO.  If both the broad and narrow
systems near $z=0.7$ are produced by ejected material, they will
evolve with time, with the broad component losing its prominence. This
raises the possibility that older versions of such systems might
appear as ``normal'' narrow intervening systems.


Could the \pg2302 broad absorption system with its velocity span of
more than 3,000~\kms be produced by some kind of large-scale
structure?  The range of velocities observed for the \pg2302 broad
absorption system is similar to that over which we reported clustering
of \Lya absorbers around extensive metal line systems (Bahcall \etal
1996).  The \pg2302 broad absorption system is also very similar to a
broad system in \QSO1038, which is most probably produced by gas
associated with some large-scale structure, perhaps a supercluster of
galaxies.  One of the two QSOs composing the ``Tololo Pair'' (see
\S~\ref{sec-intro}), \QSO1038 has a broad absorption system at
$z_{\rm{abs}}=2.08$ that is matched by similar systems in the spectra
of other QSOs in the same field.  Jakobsen \etal (1986), Sargent \&
Steidel (1987), and Dinshaw \& Impey (1996) use energy arguments to
conclude that it is unlikely for material ejected from one of these
quasars to produce the apparently correlated absorption along these
multiple lines of sight and therefore the observed broad and complex
absorption is produced by gas associated with a supercluster of
galaxies.  When the C~IV absorption complexes in these QSOs are
observed at $\sim30$~\kms resolution (Dinshaw \& Impey 1996), most are
resolved into multiple C~IV doublets, as expected if the absorption is
produced by gas associated with the halos of individual members of a
group of galaxies.  The exception is the $z_{\rm{abs}}=2.08$ system in
\QSO1038, which remains smooth and broad.  Dinshaw \& Impey (1996)
speculate that this line of sight might be passing through a cooling
flow associated with the core of a cluster and/or group of galaxies.

Like the systems in the ``Tololo Pair'', the high-ionization
absorption systems in \pg2302 span a large velocity extent and reveal
some substructure.  However, unlike the \pg2302 system, the ``Tololo
Pair'' systems all have associated low-ionization absorption lines
({\it{e.g.~}} Si~II, O~I; Sargent \& Steidel 1987), which might be
expected if the lines of sight pass through the halos and disks of
galaxies.  The broad absorption systems seen in the spectra of the
``Tololo Pair'' are imperfect analogues to the \pg2302 broad system,
and we cannot be certain that the later is caused by absorption by gas
in a cluster or supercluster of galaxies.

The observed levels of ionization of the broad system may place
interesting constraints on the ionization mechanism and metallicity of
the absorbing material. These would be aids in identifying the nature
of the absorbing gas. If collisional ionization dominates,
temperatures between 150,000--300,000 K would be consistent with our
limits on \Lya/O~VI and \Lya/N~V, but the \Lya/C~IV limit might be
more difficult to explain unless the metallicity is slightly enhanced
with respect to solar, especially if we are dealing with cooling
behind a shock front (cf. Schmutzler \& Tscharnuter 1993).
Photoionization mechanisms are consistent with our observations, but
until detections or better limits on other ions (especially C~III,
Si~III, Si~IV and \Lya) are obtained, the constraints are not very
useful (see Turnshek \etal 1996 for an example of the use of such
measurements to constrain the ionization mechanisms and metallicities
of the absorbing gas in BALQSOs).

A variety of subsequent observations might allow us to determine what
is producing the broad system in \pg2302.  Higher resolution
spectroscopy of the quasar, multi-band imaging and spectroscopy of
galaxies in the field, and X-ray observations of the field might each
provide evidence of a cluster or supercluster of galaxies at
$z\sim0.695$ and help confirm the interpretation of the absorption
features as tracers of large-scale structure.  Detection of
variability in the strength of the broad system of \pg2302 would
provide strong evidence that this system is produced by ejected
material.

\acknowledgments
This work was supported by contracts NAG5-1618, NAG5-3259, and grant
{\hbox{GO-2424.01} from the STScI, which is operated by the A.U.R.A.,
Inc., under NASA contract NAS5-26555.  We thank Digital Equipment
Corporation for providing the DEC4000 AXP Model 610 system used for
portions of this work.  Much of the Key Project software was written
by D.~Saxe and R.~Deverill.  Helpful discussions with J.~Charlton,
J.~Hill, R. Green, M.~Strauss, and A. Tanner are appreciated. We thank
C.~Foltz for the suggestion that the absorption might be produced by a
supernovae remnant.  We thank F.~Hamann, T.~Barlow, and V.~Junkkarinen
for sharing their paper on Q~2343$+$125 with us prior to its
publication.

{
}

\section*{Figure Captions}

\noindent
Figure 1.\ Ultraviolet spectra of \pg2302 obtained with the FOS of the
HST.  Each panel contains a single spectrum obtained with the
indicated grating (G270H, G190H) and the Red digicon detector.  The
G190H and G270H observations were obtained with respectively the
0.26$''$ and 0.86$''$ circular apertures.  The dotted line is the
``continuum'' fit; see \S~\ref{sec-obs} and Schneider \etal 1993.  The
lower line in each panel is the one $\sigma$ uncertainty in the flux
as a function of wavelength.

\noindent
Figure 2.\ The regions of the HST FOS spectra containing the broad
absorption by C~IV, N~V, O~VI, and the expected position for
absorption by \Lya are plotted on a velocity scale with zero velocity
corresponding to $z=0.6951$.  The location of the C~IV, N~V, O~VI, and
\Lya absorption lines from the narrower $z_{\rm{abs}}=0.7016$ system
are indicated with vertical tick marks on the figure.

\markright{}

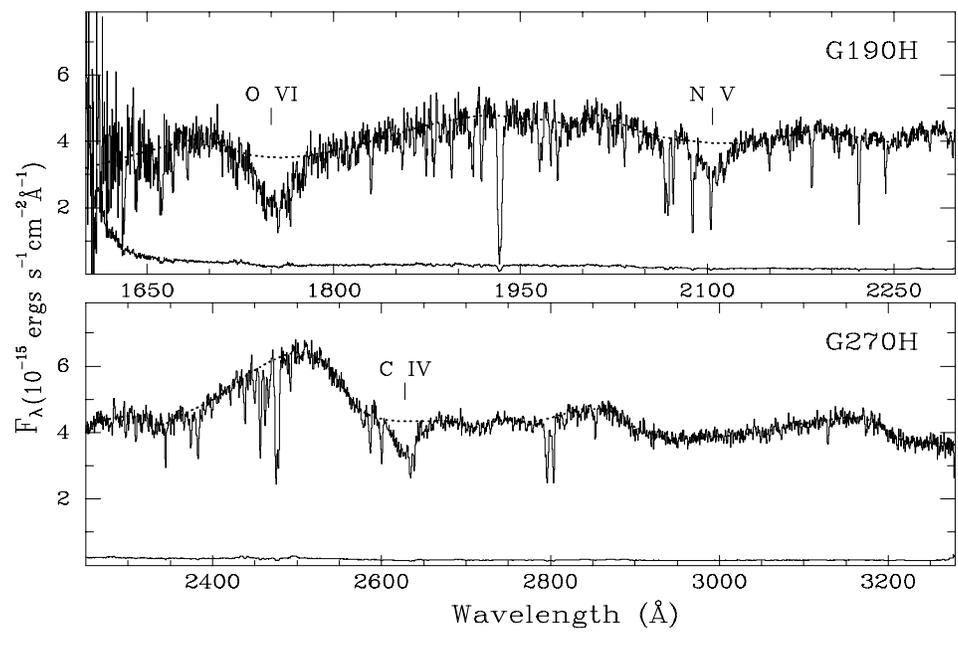
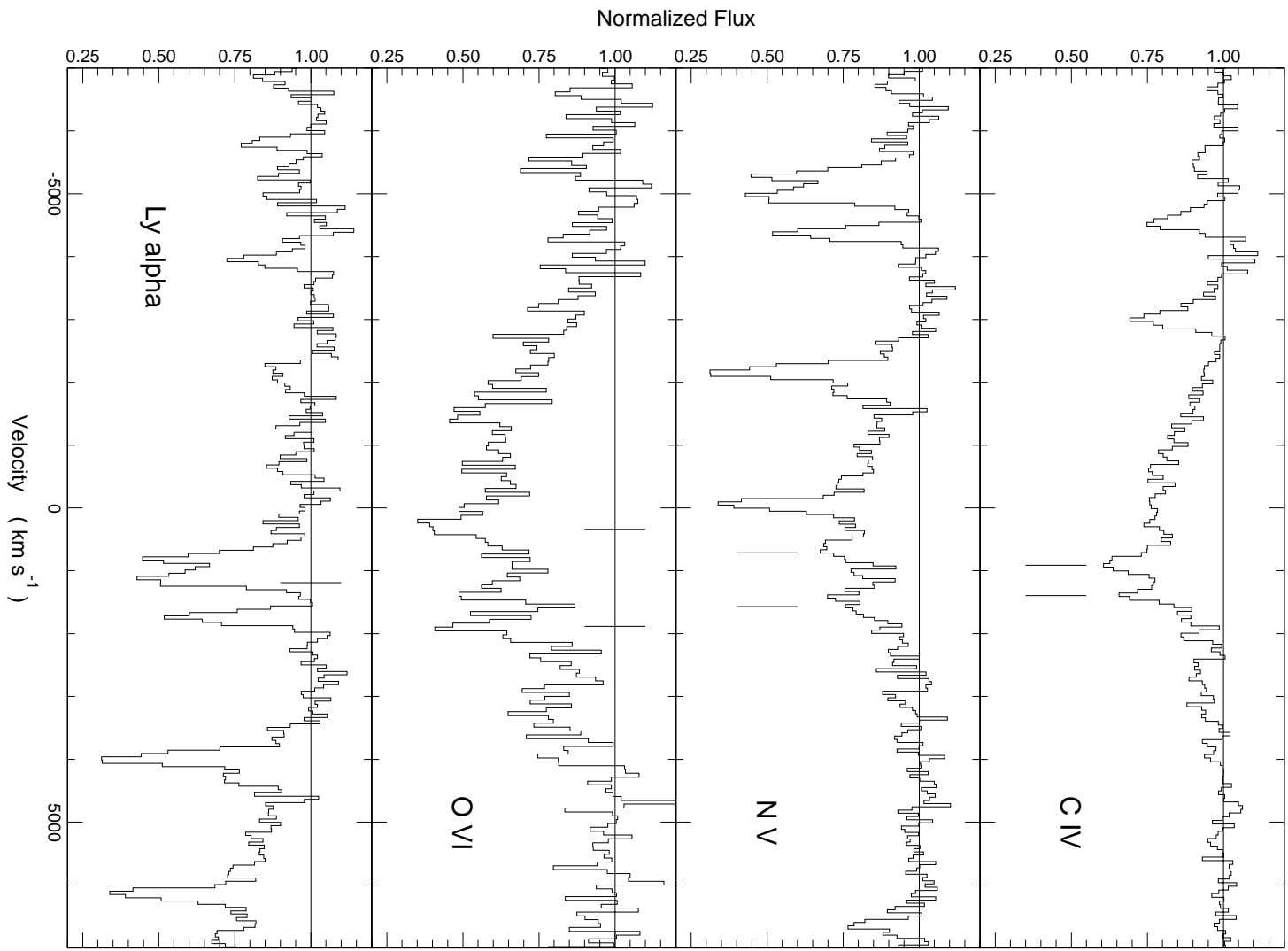
\end{document}